\documentclass[twocolumn,secnumarabic,amssymb, nobibnotes, superscriptaddress, aps, prl]{revtex4-1}
\usepackage{amsmath}
\usepackage{graphicx}
\usepackage[dvipsnames]{xcolor}

\begin{document}

\title{Hybrid Stimulated Raman Scattering - Two Plasmon Decay Instability and 3/2 Harmonic in Steep-Gradient Femtosecond Plasmas}%

\author{I. Tsymbalov}
\email{ivankrupenin2@gmail.com}
\affiliation{Faculty of Physics and International Laser Center, Lomonosov Moscow State University, 119991, Moscow, Russia \looseness=-1}
\affiliation{Institute for Nuclear Research of Russian Academy of Sciences, 117312, Moscow, Russia \looseness=-1}

\author{D. Gorlova}
\affiliation{Faculty of Physics and International Laser Center, Lomonosov Moscow State University, 119991, Moscow, Russia \looseness=-1}
\affiliation{Institute for Nuclear Research of Russian Academy of Sciences, 117312, Moscow, Russia \looseness=-1}

\author{A. Savel'ev}
\affiliation{Faculty of Physics and International Laser Center, Lomonosov Moscow State University, 119991, Moscow, Russia \looseness=-1}
\affiliation{Lebedev Physical Institute of Russian Academy of Sciences, 119991, Moscow, Russia \looseness=-1}

\date{June 2020}%

\begin{abstract}
The hybrid Stimulated Raman Scattering - Two Plasmon Decay instability had shown to play the dominant role for plasma waves excitation and energy absorption at interaction of very intense femtosecond obliquely incident $p$-polarized laser pulse (intensity $10^{17}-5\cdot10^{19}$\ W/cm$^2$) with a steep gradient ($L\sim \lambda$) plasma near the quarter critical  surface, i.e. within the typical for modern experiments conditions. The plasmons are excited as two wave packets confined near this surface with very wide $\approx \omega_0/c$ spatial spectra along its normal. Hence, phase matching conditions for 3/2 harmonic generation fulfill immediately and include new mechanism coming from high harmonics of plasma waves. The latter mechanism have been proved experimentally observing an additional 3/2 harmonic beam.

\end{abstract}
\maketitle

Parametric instabilities in a dense laser plasma have been actively studied for decades due to extensive researches on ICF and related phenomena \cite{PhysRevLett.75.4218,PhysRevLett.120.135005,PhysRevLett.124.185001}. Here inhomogeneity of plasma density $n_e$ is weak (i.e. laser wavelength $\lambda$ is much less than the plasma scalelength $L=(\frac{\partial \ln n_e}{\partial y})^{-1}$) due to the moderate intensities of nanosecond laser radiation $I\sim10^{14}-10^{16}$ W/cm$^2$ \cite{Afeyan1997,Kruer1988}. Parametric plasma waves excitation differs greatly if shorter and relativistically intense laser pulses (intensity $I>10^{18}$  W/cm$^2$) are used: (i) the increment of these instabilities is proportional to the amplitude of the laser field, (ii) amplitude of an excited plasma wave increases exponentially with time in the linear regime \cite{Kruer1988}, and (iii) preplasma, formed by an inevitable prepulse (nanosecond amplified spontaneous emission, short prepulses of various nature), or by the arbitrary long rising edge of the main pulse, is rather steep ($L \approx \lambda$) within the electron density range $n_{e} \sim 0.1-0.25n_{c}$ ($n_c$ - critical density) \cite{PhysRevLett.103.235001}, that confines parametric excitation spatially. 

Stimulated Raman Scattering (SRS) in the homogeneous plasma is intensively considered for amplification of extremely intense ultrashort laser pulses \cite{Malkin2000, PhysRevLett.120.024801} and electrons' acceleration \cite{Gordon1998}. Excitation of SRS and two-plasmon decay (TPD) instabilities by subrelativistic or relativistic femtosecond laser pulses in a long inhomogeneous plasma ($L \approx 10-100\lambda$) was studied in  \cite{Quesnel1997,Guerin1995,Gordon2002,Gordon2001,Veisz2004,Veisz2002}. 
Parametric instabilities are apparently responsible for formation of relativistic high-energy electron beams upon reflection of a powerful femtosecond laser pulse from a dense steep plasma \cite{Tsymbalov2019,Ma6980}. 

Optical emission of the 3/2 harmonic (THH) is a characteristic feature of plasma instabilities onset at a quarter-critical density \cite{Veisz2002,Veisz2004,Tarasevitch2003}. Angularly resolved emission spectra of this harmonic can be easily measured and carry information on the spatial spectra of plasma waves and their nonlinearity \cite{Downer2018}. Possible impact from the hybrid  SRS-TPD instability on the THH generation in a steep gradient plasma (SGP, $L\sim\lambda$) was pointed out earlier from experimental data \cite{Tarasevitch2003}.
However, too simple approximation of three plane waves was considered there, whereas strongly nonlinear plasma waves with a wide spatial spectrum are excited in the SGP at relativistic intensities, being confined strongly in the direction normal to the plasma surface \cite{QE}. 
Besides, the excited plasmons do not satisfy the phase matching conditions for generation of the THH immediately in this simple plane wave approximation, and wave vectors of plasmons have to change due to their propagation along the plasma gradient before the THH is emitted.

In this work we considered excitation of plasma waves by relativistic femtosecond laser pulses in the SGP with 2D PIC modeling. The mechanism of plasma waves excitation and energy absorption - the hybrid SRS-TPD instability - was proved from the ponderomotive force analysis. We showed that the phase matching conditions for generation of THH are immediately fulfilled because of the wide spatial spectra of plasma waves in the SGP. We also found out and proved experimentally the new mechanism of THH generation that considers the second harmonic of a nonlinear plasma wave.

Numerical simulations were performed using the fully relativistic 3D3V PIC code “Mandor”\cite{PhysRevLett.93.215004}, reduced to the 2D3V variant. The simulation box size was 31$\times$14 $\mu$m$^2$, spatial and temporal steps were $\lambda/100$ and $3 \times 10^{-3}$ fs respectively, total number of particles was $10^8$. The planar foil target consisted of cold electrons (initial temperature 100\ eV) and immobile ions. A $p$-polarized laser pulse with duration $\tau=$ 100 fs (FWHM) and central wavelength $\lambda=1\ \mu$m entered the simulation box with its polarization in the $xy$ plane (see Fig.1). The preplasma profile was chosen as $ n_0\propto \exp{(y/L)}$, $y$ – coordinate along the target normal with maximal density $10 n_\text c$. 

Only the TPD and SRS instabilities will remain in our study, as ions are assumed immobile at the femtosecond time scale. Let us first consider interaction of the laser pulse with $I=10^{17}$  W/cm$^2$ incident at an angle $\alpha= 60^0$ onto $L = 1.25 \lambda$ plasma gradient. This intensity is high enough to observe parametric instabilities, but hinder strong plasma turbulence, that prevents clear unveiling of instabilities development. The turning surface of the laser radiation $n_{\text{turn}}=n_c/4 $ for this angle corresponds to the maximum increment of  parametric instabilities \cite{Kruer1988}. The magnetic field of the laser pulse $H_z$ (normalized to ${mc\omega_0/e}$, where $m,e$ are an electron mass and charge, $\omega_0=2\pi c/\lambda$ - fundamental frequency) is plotted in Fig. 1a, while its spatial Fourier spectrum - in Fig. 2a. All the figures are shown at the instant when the intensity $I$ at the turning point reaches its maximum. The $x$-projection of the wave vector $k_{0x}=\frac{\omega_0}{c} \sin \alpha$ remains unchanged due to the choice of axes directions \cite{ginzburg1970propagation}, while the $y$-projection changes from the $ k_{0y}=\frac{\omega_0}{c} \cos \alpha $ for the incident radiation to the $-\frac{\omega_0}{c} \cos \alpha $ for the reflected one. These spatial harmonics form a wave packet localized near the turning surface in case of the SGP.  

Figs. 1b and 2b present electron density perturbations and their spatial spectrum. It is clear that the spatial spectrum of plasma waves looks like the spatial spectrum of the laser pump. In the following, we denote plasma waves by their $\kappa_x$ projection. Two waves with $\kappa_{1x} \approx 1.1 \frac{\omega_0}{c}$ and $\kappa_{2x} \approx -0.23\frac{\omega_0}{c}$  are amplified, while the pump wave has $k_{0x} \approx 0.87\frac{\omega_0}{c}$. The wave numbers of the first plasmon $ \kappa_{1x}$ are close to $\frac{\omega_0}{c}$, which is typical for the SRS \cite{Kruer1988} and hybrid SRS-TPD \cite{Quesnel1997} instabilities near the quarter-critical density. Next we Fourier-transformed the data in Fig. 2b back to the coordinate space using two white rectangular windows shown, thus filtering two plasmons (see Figs. 1c,d). They are localized in a narrow ($ \sim \lambda $) area in the vicinity of the quarter-critical density surface. The first plasmon (Fig. 1c) is directed along this surface, while the second one (Fig. 1d) consisting of almost a single spatial density oscillation - nearly along the density gradient. Because of the low plasma temperature, these waves do not propagate outside the quarter critical density area, where ponderomotive forces exciting them are localized.

\begin{figure*}
\includegraphics[width=\linewidth]{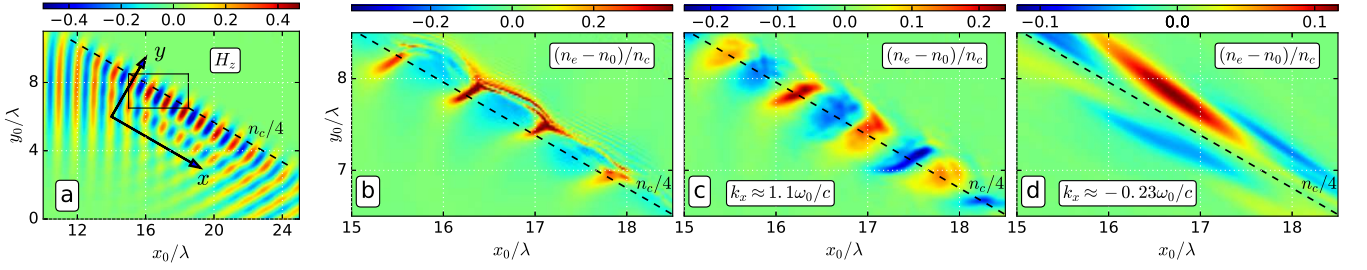}\caption{ \label{fig1} Normalized laser pulse magnetic field $H_z$ (a), electron density perturbations $(n_e-n_0)/n_c$ before (b) and after band-pass filtering in the vicinity of $k_x \approx 1.1\frac{\omega_0}{c}$ (c) and $k_x \approx -0.23\frac{\omega_0}{c}$ (d) within the rectangular windows shown in Fig. 2b. Figs.1 b-d show plasma density inside the black rectangle in Fig. 1a.}
\end{figure*}

\begin{figure*}
\includegraphics[width=\linewidth]{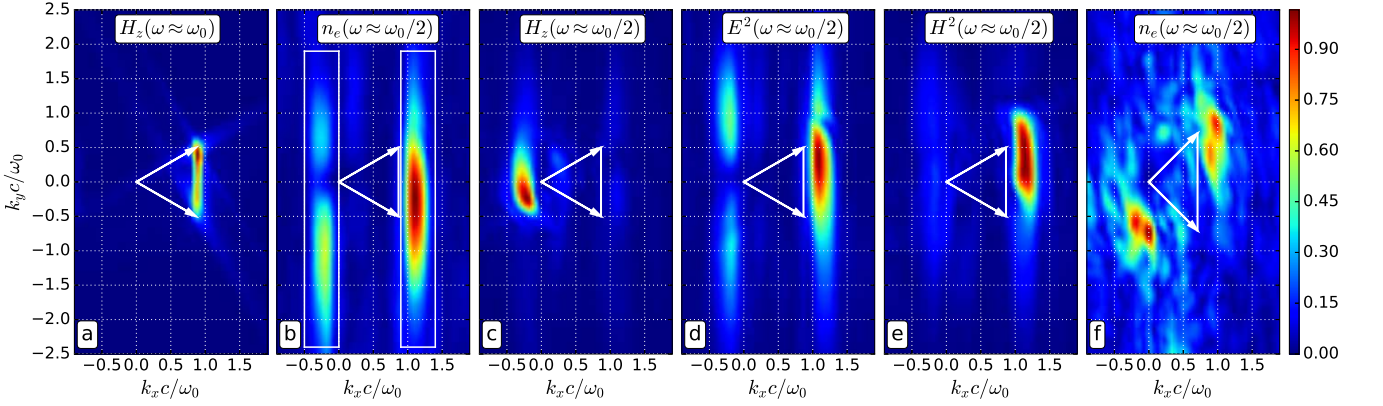}\caption{ \label{fig2} Spatial spectra (normalized to their maxima) of the laser pulse field $H_z$ (calculated from Fig. 1a) (a), plasma electron density (calculated from Fig. 1b) (b), $H_z$ component of the Stokes wave (c), quantities ${E^2}$ (d), ${H^2}$ (e) (see explanation in the text), and electron density perturbations at $I=10^{19}$ W/cm$^2$ and $\alpha=45^{\circ}$(f). Spatial spectra were obtained with preliminary Fourier bandpass filtering near the fundamental frequency $\omega_0$ for Fig. 2a and near the plasma frequency $\omega_p \approx \omega_0/2$ for Figs. 2 b--e. Arrows correspond to the wave vectors of incident and reflected laser pulses.}
\end{figure*}

Plasma waves of parametric instabilities are resonantly excited by the ponderomotive forces \cite{Mulser2010}. 
We have two longitudinal plasma waves  $\boldsymbol{E}_1(\omega_1,\boldsymbol{\kappa}_1)$ and $\boldsymbol{E}_2(\omega_2,\boldsymbol{\kappa}_2)$,  electromagnetic pump fields $\boldsymbol{E}_0(\omega_0,\boldsymbol{k}_0)$, $\boldsymbol{H}_0(\omega_0,\boldsymbol{k}_0)$ and scattered Stokes waves $\boldsymbol{E}_{\text s}(\omega_{\text s},\boldsymbol{k}_{\text s})$, $\boldsymbol{H}_{\text s}(\omega_{\text s},\boldsymbol{k}_{\text s})$. The latter waves arise when the pump wave is scattered by the plasma waves. Frequencies of the plasma waves are close to the plasma frequency $\omega_1 \approx \omega_2 \approx \omega_{\text p}$,   $\omega_{\text p} \approx \omega_0/2$ near the quarter-critical density, and $\omega_{\text s}\approx \omega_0-\omega_p\approx \omega_{\text p}$. In order to understand which components of the ponderomotive force amplify the particular plasma wave we analyzed their spatial spectra. The simplest relation between the electron density $n_e$ and the ponderomotive force $\boldsymbol{\pi_0}$ \cite{Mulser2010} in the homogeneous cold plasma can be written as $( \frac{\partial^2}{\partial t^2}+\omega_{\text p}^2 )n_e \propto \nabla \boldsymbol{\pi_0}$. This expression is linear one, hence a plasmon can be amplified by the ponderomotive force if their spatial components are collocated.

The force $\boldsymbol{\pi_0} \propto - \nabla E^2$ describes both the TPD and SRS instabilities in a first approximation  \cite{Mulser2010}. One can distinguish a part of this force $\propto -\nabla H^2$, which includes beating of electromagnetic waves only. Comparing these two quantities, one can separate contribution of the both instabilities.
We will split the ponderomotive source into the resonant to the plasma wave (i.e. having frequencies near $\omega_0/2$) and non-resonant terms: 
\begin{equation} 
\label{eq1}
	\begin{split}
	{E^2}= \boldsymbol{E}(\omega_0) \cdot  \boldsymbol{E_1}(\omega_0/2) 	+ \boldsymbol{E}(\omega_0) \cdot  \boldsymbol{E_2}(\omega_0/2) \\ + \boldsymbol{E}(\omega_0) \cdot  \boldsymbol{E_{\text s}}(\omega_0/2) + \text{nonresonant terms} 
	\end{split}
\end{equation} 
\begin{equation} 
\label{eq2}
	{H^2}= \boldsymbol{H}(\omega_0) \cdot  \boldsymbol{H_{\text s}}(\omega_0/2)+  \text{nonresonant terms} 
\end{equation} 
The resonant part of the ponderomotive force $\nabla {E^2}$ can be calculated using a bandpass filter with central frequency $\omega_0/2$ . Since the $\nabla$ operator is linear, no spatial components will be missed if we consider ${E^2}$ without the $\nabla$ operator.

Thus obtained spatial spectrum is shown in Fig. 2d. It can be seen that the spectral components of the ${E^2}$ coincide with the spatial components of plasmons in Fig. 2b, hence phase matching and amplification of both plasmons are possible. The equation \eqref{eq1} contains terms corresponding to interference of the pump wave both with plasma and electromagnetic waves, while  the resonant term in \eqref{eq2} contains only electromagnetic waves. The spatial component, which amplifies plasma waves with  $\kappa_{1x} \approx 1.1 \frac{\omega_0}{c}$ only, predominates in the ${H^2}(\omega_0/2)$ spectrum (Fig. 2e). This component comprises the backward Stokes electromagnetic wave having frequency $\omega_{\text s}\approx \omega_0/2$. Its spatial spectrum looks like the spectrum of the second plasmon with  $\kappa_{2x} \approx -0.23 \frac{\omega_0}{c}$ (Fig. 2c).  

Note, that in inhomogeneous plasma longitudinal waves may excite transverse electromagnetic waves and vice versa \cite{ginzburg1970propagation}. This is why the spatial spectrum of the ${H^2}(\omega_0/2)$ source (Fig. 2e) has weak components at $k_x\approx -0.23 \frac{\omega_0}{c}$ as the ${E^2}(\omega_0/2)$ spectrum (Fig. 2d).

Amplification of the second plasma wave with $\kappa_{2x} \approx -0.23 \frac{\omega_0}{c}$ by the electromagnetic part of the ponderomotive force (through the SRS process) is forbidden, since the corresponding $k_{\text sx} = k_{\text 0x}-\kappa_{2x}\approx1.1 \frac{\omega_0}{c}$ does not satisfy the dispersion relation $|\boldsymbol{k}_{\text s}| < 0.5 \frac{\omega_0}{c}$. Thus the overall  process can be described as the hybrid SRS-TPD instability: the $\boldsymbol{E_2}$ plasma wave is amplified by the beating of the plasma wave with the pump wave only, $\boldsymbol{E_1} \cdot  \boldsymbol{E_0}$, and corresponds to the TPD, while the $\boldsymbol{E_1}$ is amplified by the beating of the electromagnetic Stokes and plasma waves with the pump wave,  $\boldsymbol{E_{\text s}} \cdot  \boldsymbol{E_0}+\boldsymbol{E_2} \cdot  \boldsymbol{E_0}$, and is common for both the SRS and TPD processes.

The TPD instability may lead to the THH generation \cite{Tarasevitch2003,Veisz2002,PhysRevLett.52.1496} with the current  $\boldsymbol{j}_{\text {lin}}=\rho(\omega_0/2)\boldsymbol{v}_{\text osc}(\omega_0)$ being the source (here $\rho=-e(n_e-n_0)$,  $\boldsymbol{v}_{\text{osc}}(\omega_0)$ is an electron quiver velocity in the laser field). This radiation is widely used for laser plasma diagnostics \cite{BasovN.G.ZakharenkovY.A.RupasovA.A.SklizkovG.V.&Shikanov1989}.

The phase matching conditions
\begin{equation} 
\label{eq3}
\boldsymbol k_{1(2)}=\boldsymbol \kappa_{1(2)}+\boldsymbol k_0,\  \omega_{3/2}=\omega_{\text p}+\omega_0 
\end{equation}
have to be fulfilled for the frequency $\omega_{3/2}\approx 3/2\omega_0$ and the wave vectors $\boldsymbol k_{1,2}$ of the harmonic. This can not occur in a long scale length plasma ($L\gg \lambda$) immediately at the quarter critical density surface \cite{Tarasevitch2003}: wave vectors of plasmons should be tuned during their propagation in an inhomogeneous plasma. Hence, excitation of plasma waves and THH generation are separated in space.

\begin{figure*}
\includegraphics[width=\linewidth]{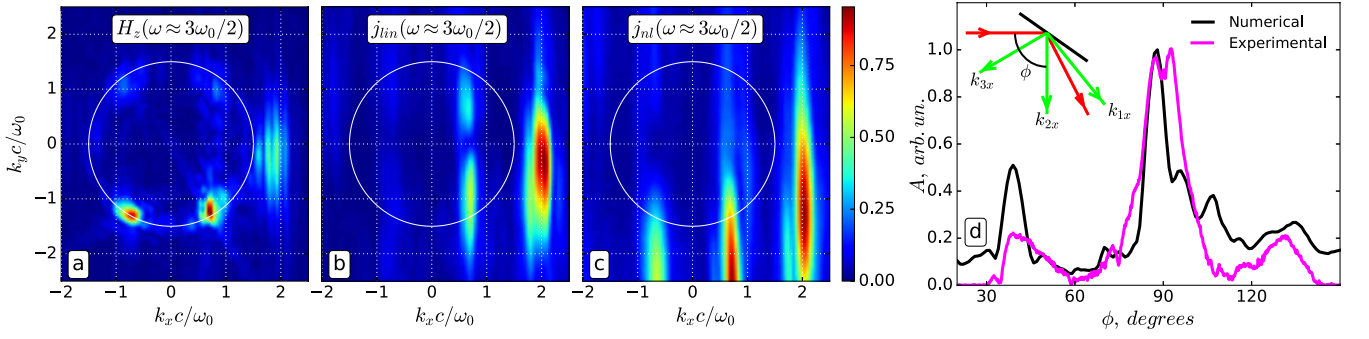}\caption{ \label{fig3} Spatial spectra (normalized to their maxima) of $H_z$ (a), current density ${j_{\text{lin}}}$ (b), current density ${j_{\text{nl}}}$ (c) after $3/2\omega_0$ bandpass filtering. Wave vectors satisfying the condition \eqref{eq3} lie on a white circle. Angular spectra of THH emission obtained in experiment (purple line) and from simulations (black line) (d). The inset shows incident and reflected laser beams (red arrows) and THH beams (green arrows).}
\end{figure*}

By the contrast, the phase matching conditions \eqref{eq3} fulfill immediately in the SGP due to the wide spatial spectrum of plasmons. This conditions can be rewritten as  ${k^2}c^2=(3/2\omega_0)^2-\omega_{\text p}^2$ that gives "radiation circle" in the $k_xk_y$ plane. Fig. 3a shows the spatial spectrum of electromagnetic waves in the simulation box filtered near the $3/2 \omega_0$ frequency. One can see two waves with $k_x \approx \pm 0.7\frac{\omega_0}{c}$ sitting exactly at the radiation circle, and the third wave at $k_{x} \approx 2\frac{\omega_0}{c}$ near the circle. Emission of the harmonic field $A$ is governed by the equation $(\frac{1}{c^2}\frac{\partial^2}{\partial t^2}- \nabla^2+\frac{\omega_{\text p}^2}{c^2})\boldsymbol{A}=\frac{4\pi}{c}\boldsymbol{j}(3/2 \omega_0)$.  Hence, spatial spectrum of the $\boldsymbol{j}(3/2 \omega_0)$ source must also intersect the radiation circle to get efficient THH emission.

The velocity $\boldsymbol{v}_{\text{osc}}(x,y,t)$ has to be calculated to obtain spatial spectrum of the linear source $\boldsymbol{j}_{\text {lin}}(3/2 \omega_0)$. It was done using the Newton's law $m_e \frac{\partial \boldsymbol{v}_{\text{osc}}}{\partial t}=-e\boldsymbol{E}$, where the electromagnetic field $\boldsymbol{E}$ was filtered near $\omega_0$. The source $\boldsymbol{j}_{\text{lin}}(3/2 \omega_0)$ was calculated using a band-pass filter at $3/2 \omega_0$ frequency. The resultant spatial spectrum is shown in Fig. 3b. There are two sources at $k_x\approx 2\frac{\omega_0}{c}$ with highest amplitude and at $k_x\approx0.64\frac{\omega_0}{c}$ with much lower amplitude. The width of the spatial spectrum along the $y$ axis  $\Delta k_{y}\sim\frac{\omega_0}{c}$ for the plasma with $L \sim \lambda$. Hence, the second source intersects radiation circle providing generation of the THH immediately after excitation of the plasmon ($k_{2x}=k_{0x}+ \kappa_{2x}\approx0.64\frac{\omega_0}{c}$ scattering process, Fig. 3a). The first  plasmon gives the $k_{1x}=k_{0x}+\kappa_{1x}\approx 2\frac{\omega_0}{c}$ scattering, but this source cannot intersect the radiation circle and corresponding THH should not appear out of plasma. Anyway, this current source is very strong and located not far from the radiation circle (Fig. 3b), while plasma is very steep. That is why the weak THH can be seen in Fig. 3a with $k_{1x}\approx 2\frac{\omega_0}{c}$.

There exists an additional spatial component in Fig. 3a at $k_{x} \approx -0.7 \frac{\omega_0}{c}$ that is absent in the source $\boldsymbol{j}_{\text{lin}}(3/2 \omega_0)$. This THH arises as a consequence of nonlinearity of plasma waves at high (relativistic) laser intensities: the instability increment amounts to $\gamma \approx 0.05\omega_0$ already at $I\approx 10^{17} \text{W/cm}^2$ near the quarter-critical density \cite{Kruer1988}. This is enough for plasma waves to become unharmonic over the duration of the laser pulse, i.e. high spatial  harmonics $\propto \exp{(-i(q\omega_p t-qk_{1(2)} x))}, q=2,3...$ appear. The new source of the THH $\boldsymbol{j}_{\text{nl}}(3/2 \omega_0)=\rho (2\omega_0/2)\boldsymbol{v}_{\text{osc}}(\omega_0/2)$ can be considered, where $\rho (2\omega_0/2)$ is the second harmonic of plasma waves and $\boldsymbol{v}_{\text{osc}}(\omega_0/2)$ - an electron quiver velocity in plasmon or Stokes wave. Fig. 3c shows spatial spectrum of this source obtained applying the proper frequency bandpass filters to the $\rho$ and $\boldsymbol{v}_{\text{osc}}$ quantities. Thus, we obtain radiation sources with wave numbers $k^{\text{nl}}_{1x}= 2\kappa_{1x}+\kappa_{2x}=2\kappa_{1x}+k_{sx}\approx1.97\frac{\omega_0}{c}$,\  $ k^{\text{nl}}_{2x}=\kappa_{1x}+2\kappa_{2x}\approx0.64\frac{\omega_0}{c}$\ and $\ k^{\text{nl}}_{3x}= 2\kappa_{2x}+\kappa_{2x}=2\kappa_{2x}+k_{sx}\approx-0.69\frac{\omega_0}{c}$. Note, that longitudinal electron oscillations generate a longitudinal current that cannot emit an electromagnetic wave in homogeneous plasma, but this is allowed in the steep gradient plasma \cite{ginzburg1970propagation}. 
The first two sources coincide with sources from $\boldsymbol{j}_{\text{lin}}(3/2 \omega_0)$, but the third source is a new one and results from plasma waves non-linearity. This nonlinear source generates the THH with  $k^{\text{nl}}_{3x}=-0.69\frac{\omega_0}{c}$ in Fig. 3a. Thus, measuring the THH spatial spectrum (which is quite easy to realise experimentally) one can get new insight into behavior of strongly nonlinear plasma waves. 

Dedicated experiments were made to verify the picture of THH generation described above. A Ti:Sapphire laser system was used with maximal pulse energy at a target $W=$50\ mJ, its shortest duration $\tau=$50\ fs and  intensity $I$ up to $5 \cdot 10^{18}$ W/cm$^2$. The $p$-polarized laser radiation was obliquely focused onto the thick molybdenum target by an off-axis parabolic mirror (the focal length $F \sim 7.5$ cm, $\alpha=60^0$). An additional pulse from the Q-switched Nd:YAG laser (maximal intensity at the target  $\sim10^{12}$\ W/cm$^2$) created pre-plasma with the controlled scalelength. A color CCD camera equipped with interference filters (to reject 800 and 1064 nm radiation) measured plasma angular emission near the $3/2\omega_0$. More detailed description of the setup, preplasma control, etc. can be found in \cite{Ivanov2017, Tsymbalov2019}. In particular plasma scalelength in the vicinity of $n_\text c$  was estimated as  $L\sim \lambda$ \cite{Ivanov2017}. In this study we chose $\tau=$ 100 fs and  $I \gtrsim 10^{17}$ W/cm$^2$ to make direct simulation-to-experiment comparison. This regime was chosen to obtain an angular distribution of the THH undisturbed by plasma turbulence that is the case at higher intensities. Fig. 3d compares experimental and numerical angular distributions of the THH. There are three main maxima at $\phi\approx 40^0, 90^0$ and $130^0$ in the both distributions. The second and third ones
correspond to the linear THH source with $k_{1x,2x}$ wave vectors. Such a quantitative agreement obviously supports our idea of immediate fulfilment of the phase matching conditions in the SGP and crucial role of wide angular spectrum of the pump and plasma waves. Note, that there are also impacts from the non-linear source with $k^{\text{nl}}_{1x,2x}$ to those two maxima. The first maximum is generated exclusively due to the non-linear THH source with $k^{\text{nl}}_{3x}$ wave vector, and this proves feasibility of the THH generation due to plasma wave non-linearity.  

\begin{figure}
\includegraphics[width=0.8\linewidth]{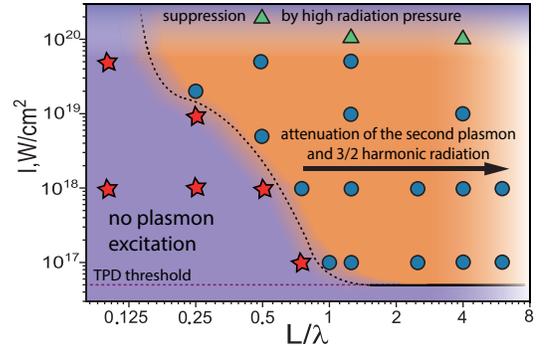}\caption{ \label{fig4} Area of the SRS-TPD instability in the $L/\lambda - I$ plane ($\tau=100$\ fs). Blue circles show simulations points where the hybrid instability is essential, red stars show points where no instabilities arise, green triangles show points, where radiation pressure becomes too high.}
\end{figure}

It is worth to establish ranges for the intensity $I$ and scalelength $L$ in which the hybrid SRS-TPD instability dominates and the picture described above stays valid. The TPD threshold intensity is $\sim 5 \cdot 10^{16}$\  W/cm$^2$ \cite{Veisz2002}. This can be considered as the lower limit. To find out the upper limit, a number of simulations were performed with intensities $10^{17}-2 \cdot 10^{20}$ W/cm$^2$ ($\tau=100$\ fs) and  different scalelengths $L=0.25-8\lambda$. Amplitude of a plasma wave reaches the wavebreaking limit \cite{doi:10.1063/1.860105} in a few laser periods at intensities $I\gtrsim 10^{18}$ W/cm$^2$, and plasma becomes strongly turbulent. Nevertheless, preliminary Fourier filtration in the frequency domain allowed us to obtain spatial spectra in which components of the hybrid instability are clearly visible (see Fig. 2f,  $I=10^{19}$ W/cm$^2$). Fig. 4 shows area where the hybrid SRS-TPD instability prevails in the $L/\lambda - I$ plane.
The hybrid instability may develop at the front of the laser pulse with intensity $I> 5 \cdot 10^{19}$ W/cm$^2$, but further the plasma profile is modified due to the high intensity of laser radiation and other processes govern plasma dynamics.
The SRS process dominates if $L \gtrsim 5\lambda$, and the second plasma wave, corresponding to the TPD is getting weaker with increase in $L/\lambda$.
The hybrid instability first occurs at $L \geq 0.5-1\ \lambda$ and subrelativistic intensities $I \sim10^{17}-10^{18}$ W/cm$^2$, however it comes into play even at $L=0.25\lambda$ in the intensity range of $ \sim 5 \cdot 10^{18}-10^{19}$ W/cm$^2$.

In conclusion, the hybrid SRS-TPD instability plays the dominant role for an ultraintense laser-plasma interaction at an oblique incidence of the $p$-polarized pulse if intensity $I$ is within $\sim 10^{17}-5\cdot 10^{19}$ W/cm$^2$ range and plasma scalelength $L$ in the $\sim 0.25-5\lambda$  range (Fig. 4). Our study was done with pulse duration $\tau=100$\ fs. Obviously, boundaries in Fig.4 change if much shorter ($\sim 10$\ fs), or much longer ($\sim 1$\ ps), pulses are considered, but the overall picture of the SRS-TPD instability stays valid. 
This instability is essential at relativistic intensities since wavebreaking of excited plasmons creates huge number of very fast relativistic electrons undergoing further acceleration and forming well collimated beams with $\sim$1 nC/J charge \cite{Tsymbalov2019, Ma6980}.

The typical spatial spectrum of plasmons in the steep gradient plasma is close to the case of homogeneous plasma in the $x$ projection with $\kappa_x\sim 1.2 \frac{\omega_0}{c}$ and $-0.2\frac{\omega_0}{c}$, but is very wide in the $y$-projection, $\Delta \kappa_y\sim \frac{\omega_0}{c}$. 
 The 3/2 harmonic is efficiently generated in the vicinity of the quarter critical surface due to immediate fulfillment of the phase matching conditions between plasmons and the pump wave in a steep gradient plasma and this is another feature of the hybrid instability in such a plasma. Hence the 3/2 harmonic is a clear signature that experimental conditions fall into the above mentioned ranges, and steep gradient preplasma is being formed. Angular spectrum of the 3/2 harmonic contains rich and valuable information on the excited plasmons. In particular, second and higher harmonics of the nonlinear plasma wave may contribute into the angular spectrum of this plasma emission.

Authors wish to thank V.Bychenkov, A.Brantov and S.Bochkarev for numerous valuable discussions and help with the Mandor PIC code. This work was done with financial support from the RFBR (grants No. 19-32-60069, 19-02-00104). Simulations were made using Moscow State University Supercomputing Facility ''Lomonosov''. D.G. acknowledges foundation for theoretical research ''Basis'' for the financial support.

\bibliographystyle{apsrev4-2}
\bibliography{references}

\end{document}